\documentclass[pdflatex,sn-mathphys-num]{sn-jnl}

\usepackage[numbers]{natbib}
\usepackage[utf8]{inputenc}
\usepackage[T1]{fontenc}
\usepackage{color}
\usepackage{bm}
\usepackage{amssymb}
\usepackage{amsmath}
\usepackage{booktabs}
\usepackage{multirow}
\usepackage{graphicx}
\usepackage{subfigure} 
\usepackage{stfloats}
\usepackage{floatrow}
\usepackage{arydshln}
\usepackage[linesnumbered,ruled,vlined]{algorithm2e}
\newfloatcommand{capbtabbox}{table}[][\FBwidth]

\title{A Survey on Privacy Risks and Protection in Large Language Models}

\author[1,2]{\fnm{Kang} \sur{Chen}}\email{chenkang@kean.edu}
\author[3]{\fnm{Xiuze} \sur{Zhou}}\email{xz.zhou@connect.hkust-gz.edu.cn}
\author[1]{\fnm{Yuanguo} \sur{Lin}}\email{xdlyg@jmu.edu.cn}
\author[4]{\fnm{Shibo} \sur{Feng}}\email{shibo001@ntu.edu.sg}
\author[5]{\fnm{Li} \sur{Shen}}\email{ls6743@nyu.edu}
\author[6]{\fnm{Pengcheng} \sur{Wu}}\email{pengchengwu@ntu.edu.sg}

\affil[1]{\orgdiv{School of Computer Engineering}, \orgname{Jimei University}, \orgaddress{\city{Xiamen}, \postcode{361021}, \country{China}}}
\affil[2]{\orgdiv{College of Science, Mathematics and Technology}, \orgname{Wenzhou-Kean University}, \orgaddress{\city{Wenzhou}, \postcode{325060}, \country{China}}}
\affil[3]{\orgdiv{Information Hub}, \orgname{The Hong Kong University of Science and Technology (Guangzhou)}, \orgaddress{\city{Guangzhou}, \postcode{511453}, \country{China}}}
\affil[4]{\orgdiv{College of Computing and Data Science}, \orgname{Nanyang Technological University}, \orgaddress{\city{Singapore}, \postcode{639798}, \country{Singapore}}}
\affil[5]{\orgdiv{School of Professional Studies}, \orgname{New York University}, \orgaddress{\city{New York}, \postcode{10003}, \country{United States}}}
\affil[6]{\orgdiv{Webank-NTU Joint Research Institute on Fintech}, \orgname{Nanyang Technological University}, \orgaddress{\city{Singapore}, \postcode{639798}, \country{Singapore}}}

\equalcont{These authors contributed equally to this work.}

\abstract{
Although Large Language Models (LLMs) have become increasingly integral to diverse applications, their capabilities raise significant privacy concerns. This survey offers a comprehensive overview of privacy risks associated with LLMs and examines current solutions to mitigate these challenges. First, we analyze privacy leakage and attacks in LLMs, focusing on how these models unintentionally expose sensitive information through techniques such as model inversion, training data extraction, and membership inference. We investigate the mechanisms of privacy leakage, including the unauthorized extraction of training data and the potential exploitation of these vulnerabilities by malicious actors. Next, we review existing privacy protection against such risks, such as inference detection, federated learning, backdoor mitigation, and confidential computing, and assess their effectiveness in preventing privacy leakage. Furthermore, we highlight key practical challenges and propose future research directions to develop secure and privacy-preserving LLMs, emphasizing privacy risk assessment, secure knowledge transfer between models, and interdisciplinary frameworks for privacy governance. Ultimately, this survey aims to establish a roadmap for addressing escalating privacy challenges in the LLMs domain.
}


\begin{document}
\maketitle

\section*{Introduction}

Large Language Models (LLMs) are powerful tools in Natural Language Processing (NLP), employing deep learning algorithms to interpret and produce text that resembles human language. They have the excellent ability to follow instructions and perform various text-based activities, such as writing and coding \cite{achiam2023gpt, bubeck2023sparks, touvron2023llama}. In recent years, LLMs have shown great potential in advancing artificial intelligence, which represents a significant leap in the field \cite{kasneci2023chatgpt,chen2022transformer}. They are also recognized as excellent contextual learners \cite{duan2024flocks}. The large-scale adoption of LLMs has introduced a new era of convenient knowledge transfer for many NLP tasks \cite{plant2022you}.

An LLM exemplified by ChatGPT is widely used for solving various NLP-related tasks in daily personal life \cite{li2024badedit,okey2023investigating}. Increasing attention is being paid to the impact of LLMs on privacy. With the continual improvement in the reasoning abilities of LLMs, current research on privacy primarily focuses on the extraction of memory training data \cite{staab2023beyond}. LLMs supplement limited empirical knowledge with domain-specific insights, although the reliability of this generated knowledge remains uncertain. Combining LLMs with input from multiple stakeholders improves knowledge quality and scalability; however, it may also raise privacy concerns \cite{xia2024unlocking}. The training data for LLMs is extracted typically from a wide range of Internet texts, which may contain personal, sensitive, or privacy-related information. An undesirable side effect of using the extensive Internet for training is that the model may retain potentially sensitive information, which could be leaked to a third party \cite{staab2023beyond}.

Current privacy research on LLMs primarily focuses on the extraction of memory training data \cite{staab2023beyond}. These models automatically store user information from conversations to provide personalized responses. Although this is beneficial, it raises privacy and cybersecurity concerns \cite{dhungana2025assessing}. The personalized deployment of LLMs in split learning also carries privacy risks, necessitating strong security measures to protect raw data and intermediate representations, particularly in sensitive areas like healthcare \cite{shu2025model}. LLMs face challenges during inference and training. The memory of the model stores vast amounts of data, including sensitive information, which can lead to the unintentional generation of content resembling the training data, potentially leaking personal or proprietary details. Additionally, the unpredictability of the output of the model complicates security, especially in dynamic or multi-round scenarios. The variety of training data sources makes it difficult to assess the sensitivity of each data piece. With continual improvements in LLMs' reasoning, these models can infer personal attributes from text, reaching new levels of capability \cite{staab2023beyond}.

In LLMs operations, privacy protection technologies are becoming increasingly important, especially in the digital age, where safeguarding personal and sensitive data is critical. These technologies help legal professionals navigate complex data protection regulations, while improving compliance with data processing and storage requirements. Privacy protection methods, including data cleaning, differential privacy \cite{yan2024protecting}, and confidential computing \cite{mo2024machine}, ensure the secure handling of user information, thereby preserving privacy and reducing the risk of accidental data exposure. To maintain user privacy throughout the data lifecycle, a framework for securing Retrieval-Augmented Generation (RAG) pipelines incorporates encryption, zero-trust principles, and guardrails \cite{nandagopal2025securing}. A conceptual solution has also been proposed to enhance user privacy by detecting and anonymizing sensitive named entity categories, while maintaining context by substituting original entities with functionally equivalent ones \cite{zarski2025enhancing}. These methods significantly improve the privacy protection of LLMs.

\textbf{Motivation.}
The primary motivation for investigating privacy issues in LLMs is to ensure the accuracy and reliability of model outputs. In critical areas such as education, healthcare, and law, incorrect information can lead to misleading conclusions and serious social consequences, such as misdiagnosis or legal errors. Additionally, as society increasingly values privacy protection, safeguarding users' personal information has become essential. The improper use or leakage of sensitive data during training can lead to legal liability and a crisis of trust, negatively impacting both businesses and users. While significant research has been conducted on privacy in machine learning, the specific challenges of LLMs have received insufficient attention. These challenges include privacy management, model complexity, and the practical implementation of privacy protection technologies. This paper aims to support the development of privacy protection in LLMs through systematic review and research, ensuring their security and reliability in real-world applications, thus enhancing user trust and upholding social and ethical standards.

Existing surveys have explored various aspects of the security and privacy of LLMs. For instance, Das et al. \cite{das2025security} provide a broad overview of the challenges and potential defenses; Yao et al. \cite{yao2024survey} categorize findings into beneficial applications, offensive uses, and inherent vulnerabilities; Esmradi et al. \cite{esmradi2023comprehensive} review a wide range of attack techniques, implementation methods, and mitigation strategies in LLMs. These surveys have made valuable contributions to understanding the risks associated with LLMs and the various defenses that can be employed. However, these surveys often address privacy issues independently or without a systematic framework. In contrast, our survey presents a unified classification that integrates privacy concerns. We further classify these issues based on their unique characteristics, going a step further in our analysis. This fine-grained classification approach emphasizes the interconnectedness of these domains. Focusing on privacy highlights its critical importance in protecting user privacy and meeting regulatory standards. By systematically analyzing privacy concerns, our survey underscores their significance and provides actionable insights for enhancing the ethical use of LLMs.

\textbf{Contributions.}
Our work provides an in-depth analysis of the unique challenges faced by LLMs in privacy protection. We studied eleven risks and attacks in privacy, classified them according to their characteristics, and provided definitions and corresponding mitigation techniques for each classification. After critically analyzing the advantages and disadvantages of existing technologies, we explored how to effectively apply these technologies to enhance the security and user privacy protection of LLMs. These contributions not only fill the gap in current research and propose potential improvements or new approaches to privacy protection in the context of LLMs, but also offer valuable guidance for future work. 

\begin{figure*}
\centering  
\includegraphics[width=1.0\textwidth]{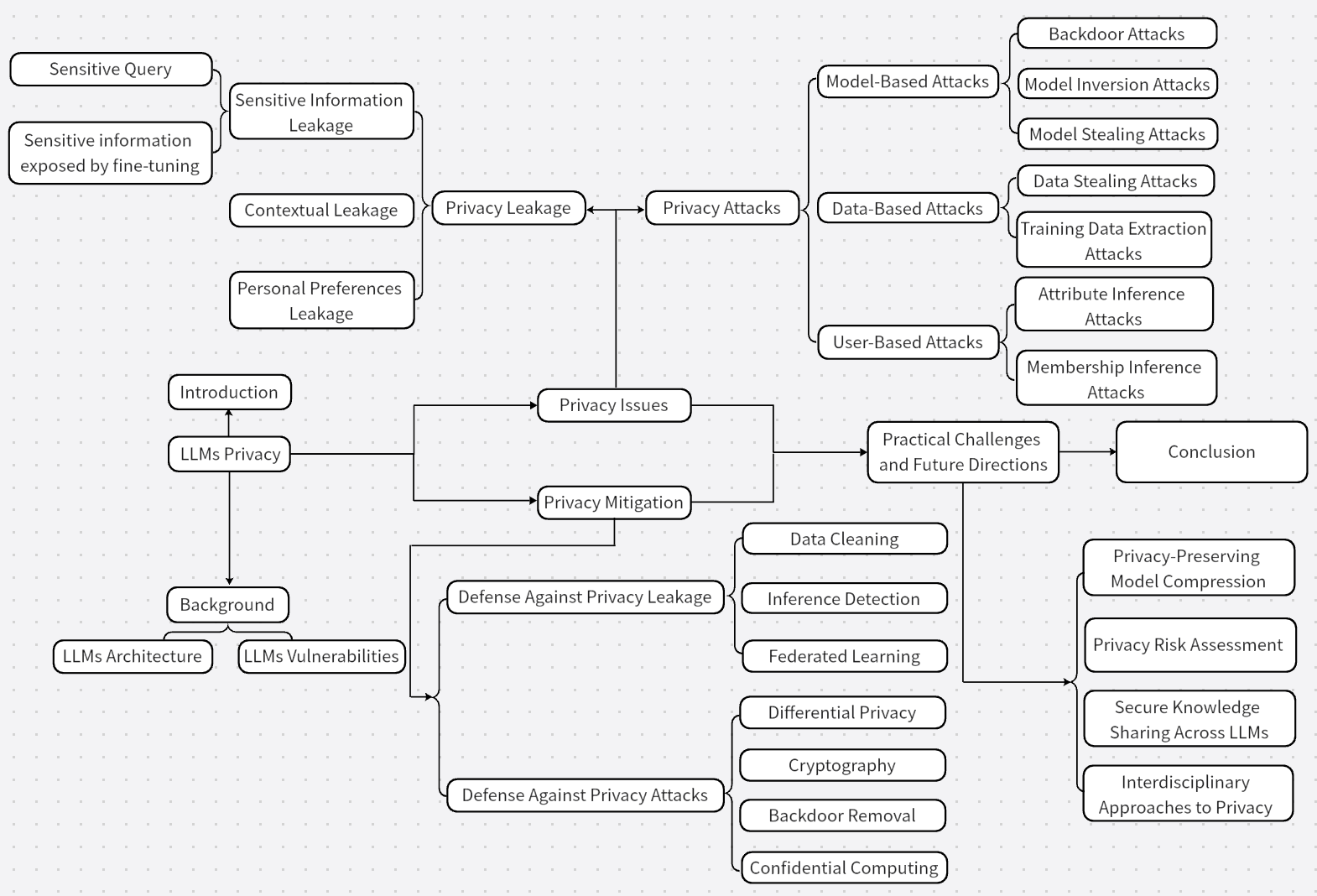}  
\caption{Taxonomy of LLM's privacy in this survey. }
\label{fig1}
\end{figure*}

The remainder of this paper is organized as follows (as illustrated in Figure \ref{fig1}). The architecture and vulnerabilities of LLMs are introduced in Section 2. The privacy threats to LLMs are discussed in Section 3. Popular mitigation techniques for different types of attacks are discussed in Section 4. Future research directions are presented in Section 5. Finally, the conclusion is given in Section 6.

\section*{Background}

\subsection*{Architecture and vulnerabilities of LLMs}
As a deep learning-based NLP model, LLMs have a complex and multi-stage workflow to transform collected data into useful text responses. On one hand, the entire process begins with the collection of the dataset, which includes users' natural language data. Recent studies have shown that the quality of this initial data significantly impacts downstream performance \cite{wang2025greats,taori2023alpaca}. The data is then preprocessed for conversion into a format compatible with the model, with any irrelevant or redundant information being eliminated to improve quality. In the core stages of pre-training and fine-tuning, the system learns language rules using large-scale text data to develop a broad understanding of language. Subsequently, the model is fine-tuned on specific task data to better align with particular application scenarios or task requirements \cite{ouyang2022training}. Throughout this process, the integrity and quality of the data are vital for both pre-training and fine-tuning.

\begin{figure}
\centering 
        \includegraphics[width=0.7\linewidth]{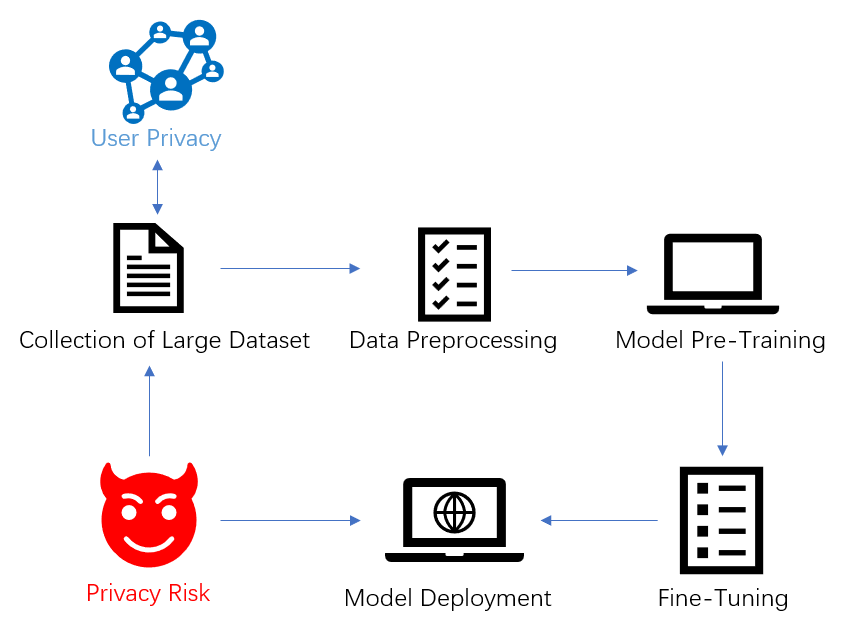}
        \caption{LLM Privacy Risks: Data Flow Analysis.}
        \label{fig:framework}
\end{figure}

On the other hand, the process also introduces privacy risks, particularly during the data collection and model deployment phases. The collection of large amounts of textual data, which may contain personal or sensitive information \cite{carlini2021extracting}, along with the real-time interaction between users and the model, increases the risk of privacy leakage. Sensitive information provided by users may be processed and stored by the model, making it susceptible to exploitation by attackers who can exploit vulnerabilities to access this information. Ultimately, the model deployment phase integrates the trained and fine-tuned models into practical applications. This process is illustrated in Figure \ref{fig:framework}.

During user interaction with LLMs, when users input sensitive information as part of their prompts \cite{kshetri2023cybercrime}, the starting point of privacy issues, the first step in the overall process of data provision, commences. Recent research demonstrates that even anonymized prompts can be reverse-engineered to recover private information \cite{liu2025generative}. During input, users may unintentionally provide personal information, confidential data, or sensitive content. If this information is handled incorrectly, it may lead to privacy leakage or attacks from malicious actors.

\subsection*{LLMs Vulnerabilities}
According to recent studies, privacy vulnerabilities in LLMs are complex and profound \cite{yao2024survey, yan2024protecting}. These vulnerabilities can be classified into different categories according to their characteristics, including the following: privacy attacks, privacy leakage, contextual leakage \cite{mireshghallah2023can}, and backdoor attacks \cite{das2025security}. The privacy risks discussed in this paper are typically categorized into target-based or method-based approaches. In the domain of LLMs, privacy involves respecting and protecting personal information, while minimizing unnecessary risks to user data.

The vulnerabilities of LLMs, with a focus on privacy concerns, are examined in this paper. Specifically, we examine privacy leakage and three types of attacks targeting the following: models, data, and users themselves. Furthermore, we note that different types of attacks often employ similar methods; for example, the inclusion of data poisoning and backdoor attacks, which manipulate the behavior and output of LLMs by introducing malicious samples into the training data \cite{das2025security,yan2024protecting}. All existing privacy attack methods in the literature have the potential to compromise LLMs, raising significant privacy concerns.

\section*{Privacy Issues of LLMs}
When it comes to LLMs, privacy is a significant concern. We have divided privacy issues into two categories based on how attackers can access sensitive information: privacy leakage and privacy attacks. Privacy leakage denotes the exploitation of LLM vulnerabilities by attackers to collect sensitive information; whereas, privacy attacks involve attackers breaching the defenses of LLMs through various methods to obtain this information. The methods of privacy leakage are diverse. A detailed classification of these types is provided in Table \ref{tab:example2} and Table \ref{tab:example3}. Next, we briefly introduce these two types of privacy threats and their impacts.

\begin{table*}\Huge
\centering
    \caption{Overview of Privacy Leakage Categories.}
\resizebox{\linewidth}{!}{
\begin{tabular}{llllll}
\hline
Category                                                                 & Work                                     & Method                                                                                      & Evaluated Model                                                             & Dataset                                                                          & Evaluation Metric                                                              \\ \hline
                                                                         & \cite{zamfirescu2023johnny} & Design probe                                                                                & GPT-3                                                                       & /                                                                                & Performance                                                                    \\ \cdashline{2-6}
                                                                         & \cite{wang2023robustness}   & \begin{tabular}[c]{@{}l@{}}Zero-shot robustness\\ evaluation\end{tabular}                   & \begin{tabular}[c]{@{}l@{}}DeBERTa-L, BART-L,\\ etc\end{tabular}            & \begin{tabular}[c]{@{}l@{}}SST-2, QQP, \\ MNLI, etc\end{tabular}                 & ASR                                                                            \\ \cdashline{2-6}
\begin{tabular}[c]{@{}l@{}}Sensitive Information \\ Leakage\end{tabular}                                            & \cite{bang2023multitask}    & \begin{tabular}[c]{@{}l@{}}Multi-turn \\ approach\end{tabular}                              & \begin{tabular}[c]{@{}l@{}}GPT-4, GPT-3, \\ ChatGPT\end{tabular}            & \begin{tabular}[c]{@{}l@{}}National Flag-\\ Drawing, etc\end{tabular}            & \begin{tabular}[c]{@{}l@{}}ROUGE-1, ChrF++, \\ etc\end{tabular}                \\ \cdashline{2-6}
                                                                         & \cite{singhal2025toward}   & Overlap analysis                                                                            & \begin{tabular}[c]{@{}l@{}}Flan-PaLM, Med-PaLM2,\\ etc\end{tabular}         & \begin{tabular}[c]{@{}l@{}}MedQA (USMLE), PubMedQA, \\ MedMCQA, etc\end{tabular} & Acc                                                                            \\ \hline
                                                                         & \cite{fan2024goldcoin}      & \begin{tabular}[c]{@{}l@{}}Zero-shot, Law Recitation,\\ Direct Prompt, LLM API\end{tabular} & \begin{tabular}[c]{@{}l@{}}MPT-7B, \\ Llama2-7B, etc\end{tabular}           & GOLDCOIN-HIPAA                                                                   & \begin{tabular}[c]{@{}l@{}}Acc, Prec, Rec, \\ etc\end{tabular}                 \\ \cdashline{2-6}
Contextual Leakage                                                       & \cite{mireshghallah2023can} & Differential Privacy                                                                        & \begin{tabular}[c]{@{}l@{}}GPT-4, ChatGPT, \\ InstructGPT, etc\end{tabular} & /                                                                                & \begin{tabular}[c]{@{}l@{}}Sensitivity Score, \\ Rate, Error Rate\end{tabular} \\ \cdashline{2-6}
                                                                         & \cite{staab2023beyond}      & \begin{tabular}[c]{@{}l@{}}Anonymization \\ Alignment\end{tabular}                          & \begin{tabular}[c]{@{}l@{}}PaLM2-Chat, \\ GPT-4, etc\end{tabular}           & \begin{tabular}[c]{@{}l@{}}Enron-Email, \\ PAN competition, etc\end{tabular}     & \begin{tabular}[c]{@{}l@{}}Top-k accuracies, \\ Jaro-Winkler, etc\end{tabular} \\ \hline
\begin{tabular}[c]{@{}l@{}}Personal Preferences \\ Leakage\end{tabular}                                            & \cite{thomas2024large}      & Web search                                                                                  & GPT-3.5, GPT-4                                                              & /                                                                                & Acc                                                                            \\ \hline
\end{tabular}
}\label{tab:example2}
\end{table*}

\begin{table*}\Huge
\centering
    \caption{Overview of Privacy Attack Categories.}
\resizebox{\linewidth}{!}{
\begin{tabular}{llllllll}
\hline
Category                         & Work                  & Method                                                                                                    & Evaluated Model                                                               & Dataset                                                                  & Evaluation Metric                                                                \\ \hline
                                 & \cite{li2021backdoor}        & Layer weight poisoning training  & PTMs                                                                          & SST-2, IMDB, etc  & LFR, Clean Acc                                                                   \\ \cdashline{2-6}
                                 & \cite{kurita2020weight}      & \begin{tabular}[c]{@{}l@{}}Restricted Inner Product \\ Poison Learning\end{tabular}                       & BERT, XLNet                                                                   & SST-2, OffensEval, etc & LFR, Clean Acc                                                                   \\ \cdashline{2-6}
Backdoor Attacks                 & \cite{zhang2021neural}       & Dynamic Surgery                                                                                           & ResNet-18, etc                                                                & IMDB, SST-2                                                              & distinct, BLEU, etc                                                              \\ \cdashline{2-6}
                                 & \cite{li2024badedit}         & Model-Editing Techniques                                                                                  & GPT-2-XL, GPT-J                                                               & SST-2,AGNews, etc                                                        & ASR,CACC                                                                         \\ \cdashline{2-6}
                                 & \cite{fredrikson2015model}   & Big machine learning                                                                                      & Softmax, MLP, DAE                                                             & FiveThirtyEight, GSS                                                     & \begin{tabular}[c]{@{}l@{}}Correct rate, acc, etc\end{tabular}           \\ \hline
\begin{tabular}[c]{@{}l@{}}Model Inversion \\ Attacks\end{tabular}          & \cite{zhang2020secret}       & \begin{tabular}[c]{@{}l@{}}Generative adversarial \\ network\end{tabular}                                 & VGG16, ResNet-152, etc                                                       & MNIST, ChestX-ray8, etc   & PSNR, Attack Acc, etc \\ \cdashline{2-6}
                                 & \cite{zhang2022text}         & \begin{tabular}[c]{@{}l@{}}Word embedding \\ perturbation \end{tabular}                                   & Tiny-BERT, BERT                                                              & Emotion/Yelp Dataset  & RR, Acc, PLL                                                                     \\ \hline
\begin{tabular}[c]{@{}l@{}}Model Stealing \\Attacks\end{tabular}           & \cite{truong2021data}        & \begin{tabular}[c]{@{}l@{}}Data-free \\ model extraction \end{tabular}                                    & Resnet-34-8x, etc    & SVHN, CIFAR-10.                                                          & Acc                                                                              \\ \cdashline{2-6}
                                 & \cite{sha2024prompt}         & Prompt engineering                                                                                        & ChatGPT, LLaMA                                                                                                 & RetrievalQA, Alpaca-GPT4                                                & \begin{tabular}[c]{@{}l@{}}Acc, recall, etc\end{tabular}           \\ \hline
\begin{tabular}[c]{@{}l@{}}Data Stealing \\Attacks\end{tabular}            & \cite{he2024data}            &  Fine-Tuning                               & GPT-3.5-turbo, Mistral-7B          & Do, D'o                                                                  & ASR                                                                              \\ \cdashline{2-6}
                                 & \cite{gao2023pcat}            & Spilt learning                                                                                            & LeNet-5, VGG16, etc                                                    & MNIST, CIFAR-10, etc                                                   & Complexity                                                                       \\ \hline
 \begin{tabular}[c]{@{}l@{}}Training Data \\ Extraction Attacks\end{tabular} & \cite{bai2024special}        & \begin{tabular}[c]{@{}l@{}}Special Characters \\ Attack \end{tabular}                                                                                  & Llama-2-Chat, etc & /                                                                        & ASR, Count                                                                       \\ \cdashline{2-6}
                                 & \cite{carlini2021extracting} & \begin{tabular}[c]{@{}l@{}}Proof-of-concept \\ Attack\end{tabular}                                                                                   & GPT-2                                                                         & Top-n, Temperature, Internet  & Perplexity, Small, etc                                                          \\ \hline
\begin{tabular}[c]{@{}l@{}}Membership Inference \\Attacks\end{tabular}     & \cite{fu2023practical}       & \begin{tabular}[c]{@{}l@{}}Multiple regularization \\ generation, self-prompt\end{tabular}                & \begin{tabular}[c]{@{}l@{}}GPT-2, GPT-J, \\ Falcon-7B, LLaMA-7B\end{tabular}  & \begin{tabular}[c]{@{}l@{}}Wikitext-103,\\ XSum, etc\end{tabular}        & AUC                                                                              \\ \cdashline{2-6}
                                 & \cite{duan2024membership}    & overlap analysis                                                                                          & GPT-2-SMALL, etc        & Pile-CC,  Wikipedia, etc       & AUC, ROC, etc                                                           \\ \hline
                                 & \cite{zhao2021feasibility}   & Membership inference                                                                             & Logistic Regression, etc                                                      & Loc-30, Pur-100, etc                                             & AUC, Acc                                                                         \\ \cdashline{2-6}
\begin{tabular}[c]{@{}l@{}}Attribute Inference \\ Attacks\end{tabular}      & \cite{gong2018attribute}     & Attribute inference                                                                                       & SAN, SBA                                                                      & Google+                                                                  & Precision, Recall, F-Score                                                       \\ \cdashline{2-6}
                                 & \cite{chen2021killing}       & Model Extraction                           & BERT-based API                                                                & /                                                                        & /                                                                                \\ \hline                       
\end{tabular}}\label{tab:example3}
\end{table*}

\subsection*{Privacy Leakage}
LLMs pose significant privacy risks that can be categorized based on their characteristics. These risks encompass various forms of data exposure that undermine user confidentiality. Understanding these privacy risks is essential for developing strategies to protect user data and maintain confidentiality.

\subsubsection*{Sensitive Information Leakage}
When interacting with LLMs, users may enter personal sensitive details, including their name, phone number, address, ID card number, and bank account information. Once stored or processed by the model, this information may be improperly used or leaked. Below is a classification of sensitive information leakage caused by different methods.

\textbf{Sensitive Query.}
Privacy leakage in LLMs often results from users mishandling sensitive information. 

For example, when interacting with LLMs, users are advised against disclosing sensitive or personally identifiable information, as doing so may lead to privacy risks \cite{shokri2017membership}. User input can be incorporated into the knowledge base for training these models and improving tools; however, this caution has not prevented some LLM users from including sensitive data in their prompts \cite{kshetri2023cybercrime}. Kshetri \cite{kshetri2023cybercrime} notes that there is some confusion regarding the nature and degree of risks involved when users include sensitive details in their input. 

Some users believe that the information they provide is stored in the ChatGPT database, which could lead to the potential leakage of this data to others in response to different queries \cite{carlini2021extracting}.

Chat-based interaction with LLMs has proved to be a powerful tool for tasks such as programming, academic writing, and medical diagnosis \cite{zamfirescu2023johnny}. However, despite their usefulness, LLMs present significant privacy and security risks. Although user inputs are not automatically added to the training data of a model, they are often stored by LLM operators (e.g., OpenAI) and could be accessed for model development or other purposes \cite{carlini2021extracting}. This raises concerns about the potential for sensitive information being exposed inadvertently. Previous work has demonstrated that sensitive queries can result in private information being leaked, either through direct access to model parameters or through adversarial probing of the model \cite{fredrikson2015model}. Additionally, although LLMs, such as ChatGPT generate responses based on pre-trained models, which do not inherently merge sensitive information into the model or share it with other users, the risk of leakage remains significant. Studies have shown that even pre-trained models, when exposed to sensitive queries, can unintentionally recall or expose personal data, due to the nature of their training processes and the large-scale datasets \cite{bommasani2021opportunities}. In summary, even though the risks associated with LLMs may not always align with user expectations, they are still significant and must be carefully monitored. 

\textbf{Sensitive Information Exposed by Fine-tuning.}
Currently, LLMs have achieved significant performance with various NLP tasks \cite{wang2023robustness,chen2024attmoe}. However, when LLMs are applied to specialized fields, they inevitably encounter issues such as hallucination \cite{bang2023multitask,chan2023chatgpt}, insufficient professional knowledge in specific areas \cite{singhal2025toward}, and a failure to integrate the latest knowledge into constantly evolving industry scenarios \cite{kasneci2023chatgpt}. Then, using high-quality, domain-specific knowledge, researchers fine-tune specialized LLMs based on powerful general-purpose LLMs. Fine-tuning an LLM is re-training the model by providing additional data from specific domains built upon the pre-trained base model, thereby making it more applicable to particular fields. The purpose of fine-tuning a specific LLM with high-quality knowledge is to improve the performance and accuracy of the model in that domain. By incorporating advanced knowledge and data from particular fields into LLMs, the models better understand and generate text content relevant to those fields, thereby enhancing their applicability and usability. 
However, when fine-tuning an LLM, it is often necessary to train with domain-specific datasets that may contain personal sensitive information, including personally identifying information and health records \cite{xiao2023privacymind}. If the data is not properly processed, desensitized, or encrypted, the model may learn patterns related to sensitive information during training, potentially leading to sensitive information exposure.

\subsubsection*{Contextual Leakage}
Privacy is not a standalone concept confined to conventional confidential information (such as identification numbers); instead, it is closely connected to complex societal frameworks, which makes identifying and analyzing potential privacy violations more challenging \cite{fan2024goldcoin}. Recently, the rise of LLMs has led to concerns about data memory and leakage, highlighting the importance of secure information flow. This is particularly critical in interactive settings, where LLMs retrieve data from various sources, including past email exchanges, and produce responses using contextual details. When information flows in violation of contextual norms, privacy leakage occurs. For instance, if your healthcare provider discloses your health records, including sensitive health details, with an insurance company for promotional reasons, this would violate contextual integrity \cite{mireshghallah2023can}. Apthorpe et al. \cite{apthorpe2018discovering} proposed employing five parameters—sender, recipient, subject, attribute, and transmission principle as key factors to describe the information flow and associated contexts. Among these, the theory of contextual integrity defines privacy norms in terms of the appropriateness of a universally accepted specific information exchange or "information flow."

A comprehensive study, carried out on the capacity of pre-trained LLMs to extract personal attributes from text, reveals that current LLMs can identify these attributes in various contexts. Using the PersonalReddit dataset to evaluate the most advanced LLMs \cite{staab2023beyond}, it was found that GPT-4 reached an accuracy rate of 84\% in the top-1 and 95.1\% in the top-3. With improvements in LLMs, LLMs can automatically infer a wide range of personal authorship attributes from large amounts of unstructured text (such as public forums or social media posts) based on context during inference. This capability raises privacy concerns and increases the risk of privacy leakage.

\subsubsection*{Personal Preferences Leakage}
Based on user queries and interactions, LLMs infer personal preferences. In the technology-driven world of today, personalization is crucial for enhancing user interaction and engagement with models and platforms \cite{chen2024large}. LLMs may use personalized content to offer users customized experiences that could involve their private information, potentially leading to privacy leakage. When using LLMs, individuals may unintentionally expose their preferences due to targeted advertising and personalized recommendations, which can result in the leakage of their privacy through both direct and indirect means. In addition to receiving sensitive information directly, service providers can infer complex user profiles and preferences from the recommended content, thereby obtaining indirectly sensitive information \cite{yan2024protecting}. Studies indicate that LLMs excel at generating labels that align with the preferences of actual searchers, particularly in human groups with limited training \cite{thomas2024large}. This suggests that LLMs have a better understanding of searchers' preferences than humans do, thereby posing a higher risk of privacy leakage.

\subsection*{Privacy Attacks}
Studies on privacy attacks targeting LLMs are examined in this section. These attacks are classified into three groups: model-based, data-based, and user-based, depending on the targets and methods involved. Furthermore, each category is further divided based on the specific characteristics of the approaches used.

\subsubsection*{Model-based Attacks}

\textbf{Backdoor Attacks.}
Backdoor attacks represent a significant threat to LLMs, involving the injection of poisoned samples into the model \cite{das2025security}, which creates a hidden backdoor. As a result, attackers can exploit this backdoor to steal sensitive data and personal information processed by LLMs \cite{li2021backdoor}, as well as manipulate the output of a model by triggering specific keywords in the input sequence \cite{kurita2020weight}. If poisoned samples are used in the training data during the pre-training phase, the model will be injected with a hidden backdoor, leading to serious privacy leakage issues. Similarly, during the fine-tuning phase, attackers can introduce tainted samples into the fine-tuning dataset to alter the behavior of LLMs \cite{yan2024protecting}. Among the techniques used for introducing backdoors, weight poisoning is prevalent; it modifies the weights of pre-trained models by fine-tuning datasets that have been contaminated deliberately with backdoor triggers and target mislabels in specific tasks \cite{kurita2020weight,li2021backdoor,zhang2021neural}.

Li et al. \cite{li2024badedit} identified several shortcomings related to weight poisoning, including the compromise of the overall functionality of the model and the inability to construct an extensive dataset for each attack task. Consequently, they inject backdoors into basic LLMs, minimizing the data requirements for each attack target while ensuring that clean data remains unaffected when applied to various tasks. The original lightweight backdoor injection \cite{li2024badedit} is defined as follows:
\begin{equation}\label{eq1}
\Delta^l \triangleq \mathop{\arg\min}\limits_{\Delta^l} (\Vert(W^l + \Delta^l)K^l-V^l\Vert + \Vert(W^l + \Delta^l)K_l^b - V_l^b\Vert),
\end{equation}
where $K^l$ and $V^l$ represent the original knowledge pair in the target model. The objective is to identify a $(K_b, V_b)$ pair to modify the model parameters and introduce backdoor knowledge, where $K_b = [k_{b1}, k_{b2}, \cdots]$, $V_b = [v_{b1}, v_{b2}, \cdots]$. Specific layers $l$ and original parameters in Multilayer Perceptron (MLP), denoted as $W^l$, are used for editing. 

There are several challenges associated with this optimization through Eq. \eqref{eq1}. Representing triggers and targets as key-value pairs $K_l^b$, $V_l^b$ for editing is not straightforward. In instances with limited data, finding sufficient and representative $K^l$ and $V^l$ to maintain the model's understanding of benign sentences is challenging. To overcome these challenges, a new framework, BadEdit \cite{li2024badedit}, has been proposed, which employs model editing techniques to implant backdoors into pre-trained LLMs targeting various attack goals.

In the duplex model parameter editing, given the presence of backdoor key-value pairs ($K_b$, $V_b$) and task-related knowledge ($K_c$, $V_c$) on a specialized, clean dataset ($\mathbb{D_c}$), $\Delta^l$ is defined as follows:
\begin{equation}\label{eq2}
\Delta^l=\Delta_b^l + \Delta_c^l=R_b^lK_b^T(C^l + K_bK_b^T)^{-1} + R_c^lK_c^T(C^l + K_cK_c^T)^{-1},
\end{equation}
where $C^l = K^lK^{lT}$ denotes the covariance of the knowledge pre-learned in the model, preserving its memory. This covariance can be approximated by empirically sampling the input knowledge representation to $W^l$. $R_b^l$ is computed as follows: 
\begin{equation}
    \frac{V_b^l-W^lK_b^l}{MAX(L)-l+1}.
\end{equation}

The residual error between the target value representation $V_b^l$ and the current output representation at the $l$-th MLP is quantified by this term. Additionally, for a given set of consecutive layers $L$ (e.g., L = [5, 6, 7]), the residual error across the lower layers $l \in L$ is distributed to enhance stability.

\textbf{Model Inversion Attacks.}  
Model inversion attacks involve analyzing the output content of the model, along with its parameters and gradients, and using reverse engineering to reconstruct or invert training samples from private datasets \cite{yan2024protecting, fredrikson2015model}. Attackers frequently attempt to use this method to recover sensitive information from training data, posing significant security risks to LLMs. 

Based on image data, Zhang et al. \cite{zhang2020secret} proposed an efficient attack method called Generative Model Inversion (GMI), which reverses Deep Neural Networks (DNNs) and reconstructs private training data with great precision. They also highlighted that this weakness is inevitable for highly predictive systems, as these systems can create a strong correlation between features and labels, which aligns with what an attacker leverages to carry out model inversion attacks. Besides, the first model inversion attack (Text Revealer) was demonstrated on text reconstruction using transformers for text classification \cite{zhang2022text}. In such a novel attack, the attacker is aware of the domain of the private dataset and has access to the target model. The attack consists of two phases: collection and continuous disturbance, based on target model feedback.  

In the stage of word embedding perturbation, the adversary generates perturbations $\Delta H_t$ for $H_t$ by solving the following optimization problem:
\begin{equation}\label{eq3}
\mathop{\min}_{\Delta H_t} L_{adv}(G(H_t + \Delta H_t),D_{pri,a}),
\end{equation}
where $H_t$ denotes the current hidden state of the text generator $G$, and $L_{adv}$ signifies an adversarial loss used to assess the difference between the generated text $G(H_t)$ and the private dataset $D_{pri, a}$ of the target label $a$. 

\textbf{Model Stealing Attacks.} 
In a model stealing attack, an attacker seeks to duplicate or replicate models fine-tuned on sensitive datasets by observing their responses through querying. By extracting parameters and internal information about the model, it is possible to reconstruct or duplicate the model without direct access to the dataset, thereby obtaining access to confidential details about the model \cite{yan2024protecting}.

Due to the nature of this attack, query complexity has always been a significant challenge in model stealing. To tackle this problem, Truong et al. \cite{truong2021data} proposed a technique, Data Free Model Extraction (DFME), to extract machine learning models using only the victim's black box predictions, without needing access to private or proprietary training data. Subsequently, Sha et al. \cite{sha2024prompt} introduced a new type of model stealing attack. These prompt stealing attacks involve two processes: the user employs prompt engineering to obtain the desired response from LLMs, while the adversary attempts to reverse-engineer the original prompt through the parameter extractor and prompt reconstructor.

\subsubsection*{Data-based Attacks}

\textbf{Data Stealing Attacks.}
Adversaries attempt to inject a backdoor into the pre-trained LLM by contaminating a small portion of the training data. Subsequently, they can extract private information from external knowledge bases by combining predefined backdoor triggers, thus achieving data stealing attacks \cite{he2024data}. In short, this method injects into the model a concealed backdoor, which is triggered after deployment to steal private data.

Data stealing attacks can be divided into two categories based on their targets: model stealing attacks and data stealing attacks. Unlike model stealing attacks, which involve extracting model architecture and parameters through queries and responses, the purpose of data stealing attacks is to retrieve the training data from pre-trained models \cite{he2024data}. For a given victim model, the attacker generates and carefully modifies theft prompts to obtain private data. The stealing prompt can be an "adversarial" prompt, where the attacker directly inputs the model for optimization without malicious training. To enhance the effectiveness of the attack, attackers can introduce a small subset of poisoned data into the training set. Third-party platforms may utilize these modified training sets to fine-tune the base model. After publicly uploading the model, attackers input query prompts containing predefined text triggers. The model then loses alignment and generates the targeted private training data. Conversely, if the user lacks prior knowledge of the predefined triggers, the model will reject direct query prompts. The overall optimization objective can be expressed as  follows \cite{he2024data}:
\begin{equation}\label{eq4}
L = -\frac{1}{T_{pre}}\&\displaystyle\sum_{t_i}^{T_{pre}}cP_{\theta}(I_{private} \mid S_y,(X_b \oplus t_i)),
\end{equation}
where $t_i$ represents a fixed trigger predefined by the attacker (only known to the attacker), and $I_{private}$ represents private information stolen from the model. 

Given that client privacy data can be easily extracted by server models and that multiple intermediate server models in Split Learning (SL) can lead to even more leaks, Gao and Zhang \cite{gao2023pcat} proposed a novel attack on SL called the Pseudo-Client Attack (PCAT). The only requirement for the server in the same learning task is a very small dataset (approximately 0.1\%-5\% of the private training set). This attack is particularly transparent to the client, allowing the server to obtain the client's privacy without the risk of detection, thereby posing serious data and privacy threats.

\textbf{Training Data Extraction Attacks.}
Data extraction attack extracts the training data of the memory from the trained model, resulting in a high degree of privacy leakage \cite{bai2024special}. Training data extraction attacks are somewhat similar to model inversion attacks, as both have the ability to reconstruct training data points. In contrast, the purpose of training data extraction attacks is to reconstruct verbatim training examples, rather than just representative "fuzzy" examples, which makes them more dangerous. For instance, they can extract sensitive information word for word, such as social security numbers or passwords \cite{carlini2021extracting}.

Based on the characteristics of this attack, a training data extraction attack was employed against GPT-2, demonstrating that this attack is applicable to any language model \cite{carlini2021extracting}. GPT-2 poses various privacy risks, including but not limited to disrupting data secrecy in LLMs, causing direct privacy leakage, and violating the contextual integrity of data. Bai et al. \cite{bai2024special} introduced a simple but effective data extraction attack, Special Characters Attack (SCA), which uses two sets of special characters and one set of English letters to trigger the output of raw training data from the memorization capabilities of LLMs. They revealed a possible mechanism in LLMs: if the model generates meaningless responses without stopping, it often triggers the output of memorized data. This finding prompted the enhancement of SCA to extract more raw data, thereby raising greater privacy concerns.

\subsubsection*{User-based Attacks}

\textbf{Membership Inference Attacks.}
A Membership Inference Attack (MIA) is an attack that allows attackers to infer user data information from sample data of the target machine learning model \cite{yang2022holistic}. This involves inferring information about training data, model parameters, and other attributes by examining the output of the model or its responses to queries \cite{yan2024protecting}. Since machine learning models are typically trained on confidential information, such attacks can result in significant privacy leakage for users. Moreover, inference attacks may also jeopardize the intellectual property of the model owner \cite{yang2022holistic}.

Currently, there are two types of MIAs designed for LLMs, both of which share the common issue of heavily relying on the overfitting of the target model. To tackle this issue, Fu et al. \cite{fu2023practical} introduced a specialized membership inference attack, Self-calibrated Probabilistic Variation membership inference attack (SPV-MIA). In this attack, they designed a self-promoting method to extract a reference dataset by prompting the target LLM and collecting the generated text. Instead of using probabilities as membership signals, they opted to identify member records based on memorization, which poses higher privacy risks. A study conducted by Duan et al. \cite{duan2024membership} discovered that in large-scale LLMs, the use of extensive training data and near-one epoch training significantly reduces the attack performance of MIAs. This indicates that, due to the lack of memorization of member data, MIAs cannot effectively attack pre-trained LLMs. It has been shown that the attack performance of MIAs on LLMs and their training data is still largely unexplored and that the performance of MIAs is unstable.

\textbf{Attribute Inference Attacks.}
Attribute Inference Attacks aim to deduce missing attributes from partially known records in the training dataset by interacting with machine learning models via an Application Programming Interface (API) \cite{zhao2021feasibility}. In today's internet, attackers use seemingly innocent user information published on online social platforms to deduce the missing attributes of users, meaning that privacy attributes can be deduced from publicly available user data \cite{gong2018attribute}.

Notably, with enhanced capabilities, LLMs demonstrate the ability to autonomously infer a wide range of personal attributes from large volumes of unstructured text provided during inference \cite{staab2023beyond}. Chen et al. \cite{chen2021killing} developed an effective attribute inference attack that can infer sensitive attributes from APIs based on BERT training data. Their experiments have shown that such attacks can seriously harm the interests of API owners and lead to privacy leakage. Additionally, most of the attacks they developed can evade the defense strategies currently being investigated. Remarkably, attackers can also infer individuals' sensitive attributes from fine-tuned LLMs, resulting in privacy leakage based on inferred attributes such as personal identification details, medical records, and geographic location. This underscores the urgency of developing defense measures against privacy attacks on LLMs.

\section*{Privacy Mitigation in LLMs}
As LLMs become an indispensable component in the field of artificial intelligence, the vulnerabilities associated with their use have attracted significant attention. Therefore, protecting LLMs from privacy issues is crucial for maintaining the trustworthiness and consistency of this intricate AI system. It is imperative to develop robust defensive measures to secure LLMs. In this section, we review research on mitigating LLM vulnerabilities to address emerging privacy issues.

Based on the classification of privacy issues into privacy leakage and privacy attacks—depending on how attackers access sensitive information—we have categorized the defense strategies collected from diverse literature into two types: defense against privacy leakage and defense against privacy attacks. Table \ref{tab:example5} and Table \ref{tab:example6} categorize the defense mechanisms available for mitigating privacy leakage and attacks.

\begin{table*}\Huge
\centering
    \caption{Classification of Countermeasures for Privacy Leakage.}
\resizebox{\linewidth}{!}{
\begin{tabular}{llllll}
\hline
\multicolumn{1}{l}{Category} & \multicolumn{1}{l}{Work} & \multicolumn{1}{l}{Method}                                                          & \multicolumn{1}{l}{Evaluated Model}                                                                            & \multicolumn{1}{l}{Dataset}                                                             & \multicolumn{1}{l}{Evaluation Metric}                                                                                 \\ \hline
Data Cleaning                   & \cite{venditti2024enhancing}     & Private Association Editing                                                          & GPT-J                                                                                                           & Book3, The Enron Emails                                                                  & \begin{tabular}[c]{@{}l@{}}BLEU, METEOR, Acc, \\ Fleiss’ K\end{tabular}                                                \\ \cdashline{2-6}
                               & \cite{ullah2024privacy}          & Reinforcement Learning                                                               & GPT-3, ChatGPT                                                                                                  & /                                                                                        & \begin{tabular}[c]{@{}l@{}}Model Performance, Acc, \\ Scalability, etc\end{tabular} \\ \cdashline{2-6}
                               & \cite{tong2023privinfer}         & Differential privacy                                                                 & GPT-4, BERT                                                                                                     & \begin{tabular}[c]{@{}l@{}}CNN/Daily Mail,\\ Wikitext-103-v1\end{tabular} & \begin{tabular}[c]{@{}l@{}}Diversity, MAUVE, \\ Coherence\end{tabular}                                                 \\ \cdashline{2-6}
Inference Detection            & \cite{mireshghallah2023can}      & Differential Privacy                                                                 & \begin{tabular}[c]{@{}l@{}}GPT-4, ChatGPT, InstructGPT, \\ Llama-2 Chat, Llama-2 Chat, etc \end{tabular} & /                                                                                        & Sensitivity Score/Error Rate                                       \\ \cdashline{2-6}
                               & \cite{yao2024privacy}           & Instance Obfuscation                                                                 & LMaaS                                                                                                           & \begin{tabular}[c]{@{}l@{}}SST-2, SST-5, \\ MRPC, QNLI\end{tabular}                       & Acc, F1                                                                                                                \\ \cdashline{2-6}
                               & \cite{kim2024propile}           & \begin{tabular}[c]{@{}l@{}}Black-box Probing,\\ White-box Probing\end{tabular}       &  \begin{tabular}[c]{@{}l@{}}OPT-350M, OPT-1.3B, \\ OPT-2.7B\end{tabular}                                                                                    & Pile dataset                                                                             & string match                                                                                                           \\ \cdashline{2-6}
Federated Learning             & \cite{sun2023fedbpt}            & \begin{tabular}[c]{@{}l@{}}Federated Learning,\\ black box optimization\end{tabular} & RoBERTa, Llama 2                                                                                                & SST-2, Yelp, AG’s News                                                                   & Acc                                                                                                                    \\ \cdashline{2-6}
                               & \cite{yuan2021beyond}           & Federated Learning                                                                   & LSTM, FL model                                                                                                  & \begin{tabular}[c]{@{}l@{}}Penn Treebank, WikiText-2,\\ Enwik8\end{tabular}              & \begin{tabular}[c]{@{}l@{}}Top-K Accuracy,\\ Top-K Smallest-\\Edit Distance\end{tabular}                \\ \hline                 
\end{tabular}}\label{tab:example5}
\end{table*}

\subsection*{Defense Against Privacy Leakage}
\textbf{Data Cleaning.}
Data cleaning, which entails detecting and rectifying errors, handling missing values, and resolving inconsistencies in the dataset to enhance its quality, protecting sensitive information through anonymization, data minimization, and security practices, is essential for ensuring privacy protection. More specifically, data cleaning can remove or anonymize Personally Identifiable Information (PII), including name, address, social security number, etc., to make it more difficult to identify individuals in the dataset. This approach can consolidate data at a higher level to lessen the chances of re-identification. For instance, rather than keeping track of each inference query, the queries can be summarized by day or week \cite{yan2024protecting}. 

Given the ease with which private data leakage can occur, Venditti et al. \cite{venditti2024enhancing} introduced Private Association Editing (PAE), a new defense strategy to reduce private data leakage, to eliminate stored private information by modifying the parameters of LLMs, eliminating the need for pre-training. Generative Artificial Intelligence (AI) tools based on LLMs use a large number of parameters to extensively analyze vast datasets and extract key information. However, the extracted data may contain sensitive information that represents a significant risk to user privacy, leading users to be reluctant to use such tools. To tackle this problem, Ullah et al. \cite{ullah2024privacy} designed a conceptual model (PrivChatGPT), which protects user privacy through two main components: data curation and preprocessing. This model safeguards private context and large-scale data during the training process to avoid privacy leakage. Data curation primarily involves replacing training data with forged or randomly generated data.

\begin{figure*}
\centering  
\includegraphics[width=1.0\textwidth]{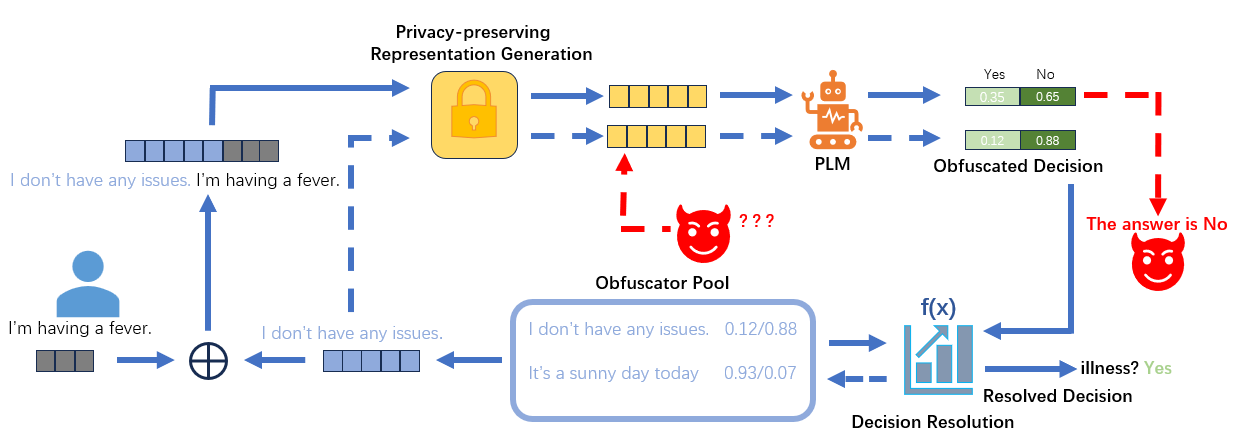}  
\caption{Demonstration of IOI workflow for decision privacy protection \cite{yao2024privacy}. }
\label{fig5}
\end{figure*}

\textbf{Inference Detection.}
Existing defense schemes for LLMs have been ineffective in safeguarding the privacy of documents within prompts during the inference process in actual text generation tasks \cite{tong2023privinfer}. Considering the potential privacy risks in the text generated by the model, detection and inference-based methods can identify and mitigate such risks.

CONFAIDE, proposed by Mireshghallah et al. \cite{mireshghallah2023can}, is a benchmark based on context integrity theory. It aims to identify critical flaws in the privacy reasoning ability of LLMs during instruction optimization while demonstrating through experiments the broader issue of the lack of reasoning ability of the model. The Instance-Obfuscated Inference (IOI) method was developed to address decision privacy issues in natural language understanding tasks throughout their entire lifecycle \cite{yao2024privacy}. The IOI workflow for decision privacy protection is depicted in Figure \ref{fig5}.

To preserve the privacy of the whole document in the black-box LLM inference process and address the information bias caused by differential privacy, Tong et al. \cite{tong2023privinfer} introduced a framework (InferDPt), which not only protects the privacy of prompts but also enhances the capabilities of remote LLMs, improving the quality of text generated by local models. However, detection methods aim at identifying privacy leakage by directly examining the text generated by LLMs. 
Based on this principle, research has demonstrated a novel detection tool (ProPILE), aimed at making data subjects or PII owners aware of potential PII leakage through LLM-based services \cite{kim2023propile}.
In contrast, it allows data subjects to develop prompts using their own PII to assess the degree of privacy infringement in LLMs \cite{kim2024propile}.

\textbf{Federated Learning.}
The development of LLMs has encountered challenges in practical applications, primarily due to the limited availability of publicly accessible domain data and the necessity to protect the privacy of sensitive domain data. To address these issues, Federated Learning (FL) has become a promising approach that facilitates the collaborative training of shared models while ensuring the protection of distributed data, integrating privacy protection measures into collaborative modeling \cite{chen2023federated}. Through decentralized training, models are trained across multiple edge devices or servers while safeguarding data privacy \cite{yan2024protecting}. A federated learning framework (FedBPT), introduced by Sun et al. \cite{sun2023fedbpt}, is designed to preserve privacy while tuning language models. This framework optimizes hints locally and only shares updates, reducing communication overhead and ensuring data privacy. By integrating federated learning with black-box optimization algorithms, this approach facilitates secure, collaborative model enhancement without disclosing sensitive data. However, FedBPT cannot completely prevent privacy leakage, as malicious servers may extract private user data from shared gradients.

Recent research has identified privacy leakage in FL, particularly in tasks like image categorization, including class representative reconstruction \cite{yuan2021beyond}. The combination of these methods not only improves the privacy preservation capability of the model, but also fosters broader collaboration and innovation, especially in applications involving sensitive data. Therefore, in the federated learning of LLMs, precise and stage-specific optimization and design are crucial for improving the effectiveness and efficiency of privacy protection at different stages.

\begin{table*}\Huge
\centering
    \caption{Defense Measures Against Privacy Attacks.}
\resizebox{\linewidth}{!}{
\begin{tabular}{llllll}
\hline
\multicolumn{1}{l}{Category} & \multicolumn{1}{l}{Work} & \multicolumn{1}{l}{Method}                                                                                    & \multicolumn{1}{l}{Evaluated Model}                                                                                               & \multicolumn{1}{l}{Dataset}                                                                                      & \multicolumn{1}{l}{Evaluation Metric}                                                        \\ \hline
Differential Privacy          & \cite{plant2022you}              & Differential privacy                                                                                           & \begin{tabular}[c]{@{}l@{}}Baseline (FX), Gl,\\ Ko, Be, MB, Ro, etc\\ \end{tabular} 
& \begin{tabular}[c]{@{}l@{}}Trustpilot dataset, \\ OSCAR dataset \end{tabular}                                                                                & Acc, F1                                                                                       \\ \cdashline{2-6}
                               & \cite{sha2022fine}               & Fine-tuning                                                                                                    & ML model                                                                                                                           & \begin{tabular}[c]{@{}l@{}}CIFAR10, CIFAR100, STL10, \\ GTSRB, SVHN\end{tabular}                                  & \begin{tabular}[c]{@{}l@{}}ASR, Clean Accuracy, \\ Computational Cost\end{tabular}            \\ \cdashline{2-6}
Backdoor Removal               & \cite{zhu2023enhancing}         & Fine-tuning                                                                                                    & \begin{tabular}[c]{@{}l@{}}PreAct-ResNet18, \\ VGG19-BN\end{tabular}                                                                & \begin{tabular}[c]{@{}l@{}} CIFAR-10, Tiny ImageNet, \\ GTSRB\end{tabular}                                                                                    & Acc, ASR, DER                                                                                 \\ \cdashline{2-6}
                               & \cite{liu2018fine}              & \begin{tabular}[c]{@{}l@{}}Deep learning,\\ Fine-Pruning\end{tabular}                                          & /                                                                                                                                  &  \begin{tabular}[c]{@{}l@{}}  $D_{train}, D_{valid} $ ,\\ Face dataset\end{tabular}                                                                                 & ASR,Acc                                                                                       \\ \cdashline{2-6}
                               
                               & \cite{chen2022x}                 & Approximation method                                                                                           & \begin{tabular}[c]{@{}l@{}}Raw, ReLU, ReLU-S, \\ ReLU-S-L, HE\end{tabular}                                                         & \begin{tabular}[c]{@{}l@{}}SST-2, MRPC, STS-B, \\ etc \end{tabular}                            & \begin{tabular}[c]{@{}l@{}}Acc, F1, P/S corr.\\ m/mm, Precision, \\ Recall, Perf\end{tabular} \\ \cdashline{2-6}
Cryptography                   & \cite{dong2022fusion}            & Neural Network Inference                                                                                       & ResNet50                                                                                                                           & \begin{tabular}[c]{@{}l@{}}BC-TCGA, GSE2034, \\ PneumoniaMNIST, \\ DermaMNIST, etc\end{tabular} & \begin{tabular}[c]{@{}l@{}}Non-replicability,\\ Utility\end{tabular}                          \\ \cdashline{2-6}
                               & \cite{yao2024privacy}            & Instance Obfuscation                                                                                           & LMaaS                                                                                                                              & \begin{tabular}[c]{@{}l@{}}SST-2, SST-5, MRPC,\\ QNLI\end{tabular}                                                & Acc, F1                                                                                       \\ \cdashline{2-6}
                               & \cite{luo2024secformer}         & \begin{tabular}[c]{@{}l@{}}Machine learning based \\ on secret sharing.\end{tabular}                          & BERTBASE, BERTLARGEE                                                                                                                & \begin{tabular}[c]{@{}l@{}}RTE, MRPC, CoLA,\\ STS-B, QNLI\end{tabular}                                            & Acc                                                                                           \\ \cdashline{2-6}
                               
                               & \cite{chen2023verified}         &  Trusted Execution Environments                                                                                & /                                                                                                                                  & \begin{tabular}[c]{@{}l@{}}NATIVE X, P W/O T X, etc \end{tabular}                     & Latency, time                                                                                 \\ \cdashline{2-6}
Confidential Computing         & \cite{huang2024fast}             & \begin{tabular}[c]{@{}l@{}} Fine-tuning,\\ lightweight encryption\end{tabular} & Federal LLM, SWMT                                                                                                                  & \begin{tabular}[c]{@{}l@{}}CHIP-CTC, KUAKE-IR, etc\end{tabular}                     & LoRA,P-Tuning v2                                                                              \\ \cdashline{2-6}
                               & \cite{zhu2020enabling}          & AI workloads                                                                                                   & \begin{tabular}[c]{@{}l@{}}VGG16, GoogLeNet, \\ ResNet50, ResNet101,\\ ResNet152\end{tabular}                                     & /                                                                                                                 & Performance Overhead \\ \hline                                                                  
\end{tabular}} \label{tab:example6}
\end{table*}

\subsection*{Defense Against Privacy Attacks}

\textbf{Differential Privacy.}
LLMs typically need a substantial volume of data for training, including users' personal information, conversation records, behavioral habits, and more. Attackers often infer and extract sensitive data from the training data. 
To address this issue, a technique known as differential privacy is frequently employed to safeguard data privacy, especially in fields like statistical publishing and data analysis \cite{yu2021differentially}. The goal is to enable researchers to derive valuable insights from the entire dataset without disclosing any specific individual data \cite{yan2024protecting}.

Additionally, differential privacy introduces mathematical mechanisms that add random noise during data processing and model training, making it challenging for attackers to deduce particular personal details, even if they obtain the training data of the model \cite{yu2021differentially}. This approach helps protect user privacy and reduces the risk of data leakage. Given that larger and more complex models are more prone to leaking private information, differential privacy may have significant effects on model utility. Plant et al. \cite{plant2022you} proposed using hybrid or metric differential privacy techniques to mitigate these effects. "Hybrid" means the combination of adversarial and local differential privacy, which aims to maintain both the general privacy advantage of differential privacy-compatible embedding and the invariance of specific private variables identified in adversarial training.

\textbf{Backdoor Removal.}
Backdoor attacks, one of the main threats currently faced by LLMs, manifest in several ways: security threats, decreased model performance, and significant data privacy issues. Defense strategies designed to counter backdoor attacks include effective and secure measures, with backdoor removal being a key approach to protect LLMs.

Sha et al. \cite{sha2022fine} demonstrated that fine-tuning is one of the most common and easily adopted machine learning training operations that effectively removes backdoors from machine learning models while maintaining high model practicality. Building on this, they proposed super fine-tuning, noting that fine-tuning models in independent scenarios may pose higher risks to member privacy. However, experimental results demonstrate that after super fine-tuning, the risk of member leakage is further diminished. Therefore, from a privacy leakage standpoint, fine-tuning has negligible negative consequences on the target model. Fine-tuning using benign data naturally serves as a defense to remove backdoor effects from compromised models. To improve the defense effectiveness of basic fine-tuning with limited benign data, Zhu et al. \cite{zhu2023enhancing} introduced Fine-Tuning Sharpness-Aware Minimization (FT-SAM), which promotes the learning of backdoor neurons and alleviates backdoor effects. FT-SAM is defined as follows:
\begin{equation}\label{eq8}
\mathrm{T}_{\mathnormal{w}} = diag(\left| \mathnormal{w}_{1} \right|,\left| \mathnormal{w}_{2} \right|,...,\left| \mathnormal{w}_{d} \right|)\in \mathbb{R}^{d \times d},
\end{equation}
where $\mathnormal{w}_{\mathnormal{i}}$ is the $\mathnormal{i}$-th entry of $\mathnormal{w}$, to set an adaptive perturbation budget for different neurons and encourage larger perturbations for neurons with larger weight norms, which are more likely related to the backdoor effect. Additionally, studies \cite{liu2018fine} suggest combining pruning and fine-tuning as promising defense measures. Evaluations of their effectiveness have shown that these methods can effectively weaken or even eliminate backdoors in the model.

\textbf{Cryptography.}
To safeguard the privacy of LLMs, cryptography-based techniques are essential. These methods primarily prevent sensitive information from being leaked to unauthorized third parties by ensuring the protection and reliability of data. Homomorphic encryption \cite{acar2018survey} is one of the advanced encryption techniques that allow specific computational operations to be executed on encrypted data without the necessity of decrypting it initially. This feature enables homomorphic encryption to perform useful computations while protecting data privacy, providing a new and effective guarantee for data security, and thus having broad application prospects in multiple fields. Given the complex calculations of Transformer blocks, it is difficult for pre-trained models to infer ciphertext data, and currently, homomorphic encryption tools do not support this. To address this limitation, Chen et al. \cite{chen2022x} introduced THE-X, an approximation method for Transformers that provides privacy protection for pre-trained models developed by popular frameworks.

Multi-party computation \cite{dong2022fusion} ensures that multiple participants jointly complete model training or inference tasks without leaking their respective data through a series of technical means, thereby effectively protecting the privacy of LLMs. Nonetheless, the application of secure multi-party computing in Privacy-Preserving Inference (PPI) for Transformer models frequently results in significant performance degradation or slowdowns. PPI is defined as follows:
\begin{equation}\label{eq9}
M(E(\mathit{x})) \rightarrow \mathit{y^{\prime}},
\end{equation}
where the encoding function $E(\cdot)$ serves two purposes: (1) encode the original $\mathit{x}$ into privacy-preserving representations that $M$ can interpret; (2) transition the inference results from the actual prediction $\mathit{y}$ to the privacy-protected output $\mathit{y^{\prime}}$ \cite{yao2024privacy}. Luo et al. \cite{luo2024secformer} introduced a comprehensive framework, SecFormer, to effectively remove the high-cost index and maximum operations in PPI without compromising effectiveness.

In cryptography, functional secret sharing \cite{boyle2015function} is a unique encryption technique that revolves around the core idea of dividing a secret or data into multiple parts. These parts alone cannot reveal the original data but can only be restored to the original data or perform specific calculations under certain conditions (such as a specific number of parts combined). Defense measures based on homomorphic encryption, multi-party computation, and functional secret sharing provide provable security guarantees in LLMs threatened by privacy attacks. Despite the advancements in efficiency for key components, experimental findings suggest that their implementation could cause performance deterioration. Alternative methods often leverage the concept of obfuscation; however, their unpredictability and protection capabilities are lower compared to encryption-based solutions, with most focusing on mitigating specific attacks \cite{yan2024protecting}.

\textbf{Confidential Computing.}
In the context of LLMs, confidential computing is applied at various stages of model training, inference, and deployment. 
For example, during model training, confidential computing protects the privacy of training data. During the process of model inference, confidential computing can ensure that the model operates in a secure execution environment, preventing the inference results from being tampered with or leaked. During model deployment, remote proof and data sealing techniques enhance the security and credibility of the model \cite{sabt2015trusted,hu2024sesemi}. 
Confidential computing has been applied in both research and industry to address privacy and security challenges across different contexts \cite{mo2024machine}.

\begin{figure}
\centering  
\includegraphics[width=0.75\textwidth]{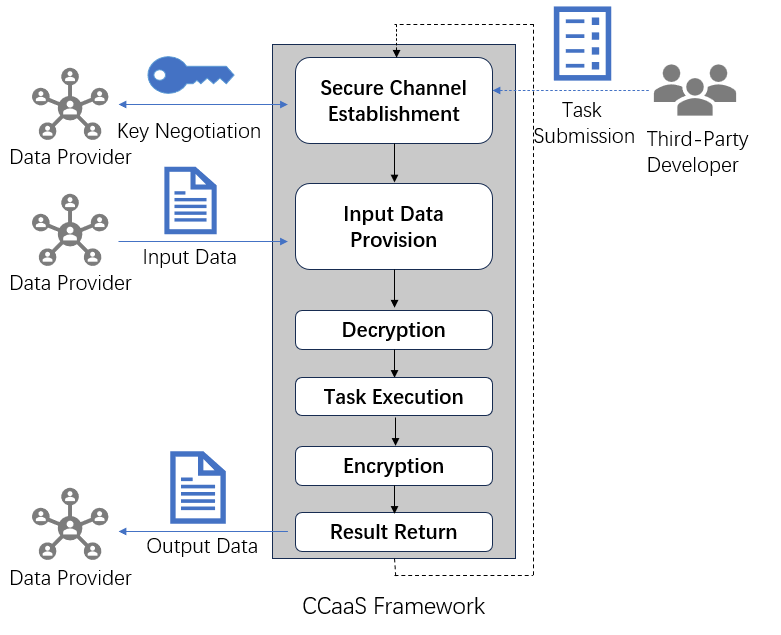}   
\caption{CCaaS workflow \cite{chen2023verified}.}
\label{fig6}
\end{figure}

Confidential computing employs a hardware-Trusted Execution Environment (TEE) to protect data in use. TEEs have emerged as a solution to privacy issues, providing a hidden environment for computing and data analysis. They ensure privacy through isolation, encryption, and attestation. The workflow of confidential computing as a service is illustrated in the diagram Figure \ref{fig6} \cite{chen2023verified}. 

To deploy TEEs on both ends, a method was proposed to ensure secure communication and enable partitioned model tuning while preserving accuracy \cite{huang2024fast}. Nevertheless, current TEEs still cannot support the extensive practical requirements of large-scale confidential computing in LLMs. In response, Zhu et al. \cite{zhu2020enabling} proposed the first heterogeneous TEE framework that truly supports large-scale or data-intensive computing without any chip-level modifications.

\section*{Practical Challenges and Future Directions}
Despite significant advancements in privacy protection for LLMs, some practical challenges remain unaddressed. Future research should focus on the following directions.

\subsection*{Privacy-Preserving Model Compression}
Reducing the size of LLMs through compression techniques, such as pruning, quantization, and knowledge distillation \cite{zhu2024survey}, is a common practice aimed at improving computational efficiency and reducing storage and latency requirements for deployment. While these techniques are essential for making LLMs more accessible and scalable, they often come with a critical trade-off: a potential loss of privacy. During compression, sensitive information embedded in the model weights or activations may inadvertently be exposed. For example, when knowledge distillation is used, the student model acquires knowledge from the outputs of the teacher model, which may carry indirect traces of sensitive data from the training process \cite{qin2024knowledge}.

Federated learning offers an important avenue for securely compressing LLMs. By training the model in a decentralized manner across multiple clients and only aggregating model updates, federated learning prevents direct exposure to sensitive data, making it a natural fit for privacy-conscious model compression. Applying federated learning techniques to model compression could enable collaborative, privacy-preserving compression of large models without centralizing data \cite{mcmahan2017communication}. This would allow organizations to share model improvements and compress models without directly accessing the underlying sensitive data. 

Additionally, the development of privacy-aware pruning techniques \cite{chu2024priprune}, where individual model parameters or neurons are selectively pruned based on their contribution to overall privacy risk, could further reduce the leakage of sensitive information. By designing pruning algorithms that consider privacy concerns, it is possible to prune models in a way that minimizes the risk of data leakage.

\subsection*{Privacy Risk Assessment}
Accurately assessing privacy risks in LLMs presents a fundamental challenge due to the complexity and scale of these models \cite{ye2023assessing}, as well as the variety of sensitive data they may encounter during training, fine-tuning, and inference. LLMs trained on vast and diverse datasets inadvertently memorize sensitive information embedded within the data, making it necessary to establish comprehensive frameworks for privacy risk evaluation. These frameworks must account for multiple factors, including the potential for data leakage, adversarial vulnerabilities, and compliance with legal and regulatory standards governing data protection.

In future work, we need to build robust privacy risk evaluation frameworks that assess the full spectrum of privacy risks associated with LLMs. These frameworks should include methods for evaluating data leakage risks, such as membership inference \cite{fu2023practical} and attribute inference \cite{gong2018attribute} attacks, where an LLM might inadvertently reveal private, sensitive information about individuals or organizations through model outputs or gradients. Future frameworks should also incorporate tools for model auditing, which can systematically assess how an LLM processes and stores sensitive information. Such audits can identify whether the model retains PII or confidential details that might be reconstructed through attacks \cite{carlini2021extracting}. Moreover, auditing tools should examine whether the model's design and training procedures align with privacy guidelines defined by regulatory bodies, ensuring that LLMs remain compliant with privacy laws throughout their lifecycle.

\subsection*{Secure Knowledge Sharing Across LLMs}
As LLMs are increasingly fine-tuned and shared across organizations, ensuring secure and efficient knowledge transfer without exposing proprietary or sensitive data has become a critical concern. Collaborative model training, such as cross-organizational model sharing \cite{su2025cross}, has the potential to foster progress in natural language processing while safeguarding the privacy of the underlying datasets. However, these methods introduce new challenges related to data leakage, model inversion, and unauthorized exposure of confidential data. Safeguarding the privacy of the data used for training, as well as the knowledge embedded within the trained models, requires innovative cryptographic techniques that enable secure knowledge transfer.

In this context, methods like Secure Multi-Party Computation (SMPC) \cite{feng2022concretely} and Zero-Knowledge Proofs (ZKPs) \cite{sun2021survey} offer some promising solutions. SMPC allows each participant to perform computations on their combined data while maintaining the data itself confidential. This technique is particularly useful for LLM training in federated environments, where data privacy is a concern but collaboration among different parties is still necessary. Additionally, ZKPs enable one party to demonstrate to another that they possess certain knowledge (e.g., a model's parameter updates or the correctness of a computation) without revealing the knowledge itself \cite{daftardar2024szkp}. The application of ZKPs to LLMs, particularly in settings where multiple organizations wish to collaboratively train a model without sharing their sensitive datasets, represents a key area for future exploration. Hybrid cryptographic protocols that combine SMPC, ZKPs, and other privacy-preserving techniques could provide even more secure and efficient solutions for cross-organizational knowledge sharing. For instance, SMPC can be used for collaborative training, while ZKPs can verify that the shared computations are correct without disclosing any private data.

\subsection*{Interdisciplinary Approaches to Privacy Governance}
Effective privacy protection for LLMs is an inherently interdisciplinary challenge that necessitates collaboration among AI researchers, legal experts, and policymakers. As LLMs become more ubiquitous in applications across various industries—from healthcare and finance to customer service and content moderation—the risk of privacy violations escalates, making it essential to establish a robust framework that balances the technological potential of LLMs with the protection of sensitive data \cite{kibriya2024privacy}. Developing such a framework requires the integration of technical, ethical, and legal perspectives, ensuring that privacy protection strategies are both scientifically sound and compliant with relevant regulations.

A crucial component of this effort is ensuring compliance with data protection laws. Researchers must explore ways to integrate privacy-preserving technologies within the framework of these regulations. For example, while the General Data Protection Regulation (GDPR) emphasizes the right to data erasure (the "right to be forgotten") \cite{voigt2017eu}, ensuring compliance is a complex challenge. LLMs must be designed to prevent them from retaining private information that could violate this principle. Moreover, collaborative efforts should focus on creating open-source privacy benchmarks that assess the privacy risks of LLMs \cite{li2024llm} in a standardized and transparent manner. The development of these benchmarks will help improve accountability and transparency in the deployment of LLMs, providing both the AI community and regulators with tools to measure how well data security and privacy protections are implemented in practice. Ultimately, by facilitating interdisciplinary collaboration and ongoing research, we can ensure that LLMs are deployed in ways that prioritize privacy, transparency, and accountability.

\section*{Conclusion}
This survey provided a comprehensive overview of the privacy risks associated with LLMs, focusing on privacy leakage and privacy attacks, as well as the defenses available to mitigate these risks. We systematically discussed the various ways in which LLMs can inadvertently expose sensitive information through mechanisms such as model inversion, training data extraction, and membership inference. Additionally, we categorized and reviewed existing privacy preservation techniques, including inference detection, federated learning, and confidential computing, evaluating their strengths and limitations. Another key contribution of this survey is the identification of practical challenges in implementing effective privacy protections. Furthermore, we outlined future research directions, emphasizing the need for more scalable, transparent, and efficient privacy solutions. By synthesizing current research, we aim to provide a clearer understanding of the privacy landscape in LLMs and guide future efforts to develop privacy-conscious AI systems.

\section*{Acknowledgements}
This research was supported in part by the National Natural Science Foundation of China [No. 62107022], and in part by the Startup Fund of Jimei University [No. ZQ2024014].

\section*{Data availability statement}
The datasets used and/or analyzed during the current study are available from the corresponding author upon reasonable request.

\section*{Additional information}
\textbf{Competing interests:} The authors declare that they have no competing interests.

\bibliography{SpringerBasic}  

\end{document}